\documentstyle[12pt]{article}
\begin{document}
\makeatother
\renewcommand{\theequation}{\thesection.\arabic{equation}}

\date{November 1993}
\title
{Scaling and Density of Lee--Yang Zeroes in the four
Dimensional Ising Model
\thanks{Supported by Fonds zur F\"orderung der
Wissenschaftlichen Forschung in \"Osterreich, project P7849.} }
\author
{\bf R. Kenna and C.B. Lang \\ \\
Institut f\"ur Theoretische Physik,\\
Universit\"at Graz, A-8010 Graz, AUSTRIA}
\maketitle
\begin{abstract}
The scaling behaviour of the edge of the Lee--Yang zeroes in the four
dimensional Ising model is analyzed. This model is believed to belong
to the same universality class as the $\phi^4_4$ model which plays a
central role in relativistic quantum field theory.  While in the
thermodynamic limit the scaling of the Yang--Lee edge is not modified
by multiplicative logarithmic corrections, such corrections are
manifest in the corresponding finite--size formulae. The asymptotic
form for the density of zeroes which recovers the scaling behaviour of
the susceptibility and the specific heat in the thermodynamic limit is
found to exhibit logarithmic corrections too. The density of zeroes for
a finite--size system is examined both analytically and numerically.
\end{abstract}

\vspace*{2.0cm}
\noindent
\parbox{10cm}{PACS number(s): 05.50.+q,02.70.+d,05.70.Fh}

\newpage

\section{Zeroes of the Partition Function}
\setcounter{equation}{0}

The 4D Ising model is believed to belong to the same universality class as the
$\phi^4$ model which plays a central role in relativistic quantum
field theory. The grand canonical partition function for the Ising model
(which corresponds to the vacuum to vacuum transition amplitude in $\phi^4$
theory) in the presence of an external magnetic field $H$ is
\begin{equation}
 Z_N = \frac{1}{\cal{N}}
\sum_{\{\phi_i\}} e^{-\kappa
 \left(
 -J\sum_{\langle ij \rangle }\phi_i\phi_j-H\sum_i\phi_i
 \right)} \quad,
\label{gndcanIsing}
\end{equation}
where $\kappa = (k_BT)^{-1}$ is the inverse of the Boltzmann constant times
the
temperature, $J$ is a coupling constant representing the strength of
the intersite interaction (set to unity in the following) and $N$ represents
the total number of sites on the lattice. The value, $\phi_i$, of the spin
at site $i$, is restricted to $\pm 1$. In its generic form, only nearest
neighbour interactions are considered and such a link is represented
by ${\langle i,j \rangle}$. The sum runs over all $\cal{N}$
possible configurations of the spin field, and the normalization ensures
$Z_N=1$ when $\kappa = 0$. The partition function may be expressed as
\begin{equation}
 Z_N = \sum_{M=-N}^N
 \sum_{S=-dN}^{dN}
 \rho\left(S,M\right)e^{\kappa S}e^{h M}
 \quad ,
\label{gndcanIsing1}
\end{equation}
in which $d$ is the dimensionality of the system, $h=\kappa H$ is the
reduced external magnetic field, and
\begin{equation}
 S = \sum_{\langle i,j \rangle} \phi_i \phi_{j}
\quad , \quad M = \sum_{i=1}^N \phi_i \quad ,
\label{paperI3.3}
\end{equation}
are the configuration energy and magnetization.  In $S$  the sum is
over the $dN$ nearest neighbours or links of a periodic lattice.  The
spectral density $\rho (S,M)$ denotes the relative weight of
configurations having given values of $S$ and $M$. In the absence of an
odd external field a second order phase transition occurs at a critical
value $\kappa_c$ of $\kappa$. The reduced temperature
\begin{equation}
 t = \frac{\kappa_c - \kappa}{\kappa_c}
\end{equation}
is a measure of the distance away from criticality. The partition function
$Z_N$ can be written as an $N^{\rm{th}}$ degree polynomial in the
fugacity $z$ defined by
\begin{equation}
 z = e^{-2h} \quad,
\label{paperIII2}
\end{equation}
as
\begin{equation}
Z_N(t,z) = z^{-\frac{N}{2}}
 \sum_{k=0}^{N} \rho_{k}(t) z^k \quad,
\label{polynomial}
\end{equation}
in which
\begin{equation}
 \rho_k(t) = \sum_{S=-dN}^{dN}\rho(S,N-2k)e^{\kappa S}
\end{equation}
is an integrated density.

That the partition function (\ref{polynomial}) is analytic for finite
$N$ establishes that no phase transition can occur in a finite--size
system. However as $N$ is allowed to approach infinity, phase
transitions which manifest themselves as points of non--analyticity can
and do occur.  In 1952 Lee and Yang \cite{LY} showed that the study of
the onset of criticality is equivalent to that of the scaling behaviour
of the zeroes of the partition function.  For a finite system, or in
the thermodynamic limit but in the symmetric phase ($t>0$), the zeroes
in $H$ are strictly complex and the free energy is analytic in a
non--vanishing neighbourhood of the real axis. As criticality is
approached ($N \rightarrow \infty$, $t \rightarrow 0$) the Lee--Yang
zeroes pinch the real $H$ axis, precipitating a phase transition. The
Lee--Yang theorem states that for the Ising model these zeroes lie on
the unit circle in the complex fugacity plane (the imaginary axis in
the complex external field plane). This theorem holds independent of
the size, dimension and structure of the lattice.

In the forty years since the ideas of Lee and Yang were presented,
there has been continual interest in this approach to the problem of
phase transitions. Analytical progress has included alternative and
modified proofs of the original circle theorem, extensions of the
result to other systems, as well as theorems proving that no zeroes can
exist in certain regions (see \cite{thesis} for a review).

It has been shown  rigorously that for isotropic nearest neighbour
interactions, and for $t$ sufficiently positive (the symmetric phase),
there exists a region around $H=0$ which is free from zeroes
\cite{GaMSRo67}.  This means there exists a gap $|{\rm{Im}}H| < H_1(t)$
where the density of zeroes is zero. The free energy is analytic in $H$
in the gap and no phase transition can occur (as a function of $H$).
The point $H=iH_1(t)$, which is a branch point of the partition
function, is called the Yang--Lee edge \cite{Fi78}.  One expects that
this property (the existence of a gap) holds in fact for all $t > 0$.

Early numerical work on Lee--Yang zeroes involved the exact calculation
of the density of states $\rho(S,M)$ and was therefore restricted to
very small lattice volumes \cite{KaAbYa71}. In the 1980's, Monte Carlo
histogram approximations to the density of states \cite{FaMa81}, the
invention of cluster algorithms \cite{SwWa87,Wo89} and multi-histogram
methods \cite{Bhanot,FeSw88,AlBeVi90} all provided boosts to the
numerical approach. Nonetheless, numerical studies are necessarily
limited to finite volume.

It is, however, the infinite volume limit which is of primary
interest.  An important analytical breakthrough which related such
numerical analyses in finite volume to the thermodynamic limit came
when Itzykson, Pearson and Zuber \cite{IPZ} connected the concept of
partition function zeroes to the renormalization group and thereby
formulated a finite--size scaling theory for these zeroes.  Their work
applies to dimensions of three or less. This was later extended to
dimensions of five or more in \cite{GlPrSc87}.  The upper critical
dimension of the Ising or $\phi^4$ universality class is $d = 4$. Above
this the scaling behaviour of the thermodynamic functions simplifies
and the critical exponents are exactly those of the mean field theory.
Below four dimensions the scaling behaviour is of a power--law type. At
$d=4$ the mean field power--law scaling behaviour is modified by
multiplicative logarithmic corrections --- a circumstance intimately
related to the expected triviality of the theory
\cite{AiGr83,ADCCaFr,HaTa87}.  Logarithmic corrections to the
finite--size scaling of partition function zeroes in four dimensions
have recently been identified from a perturbative renormalization group
analysis backed up by a high precision numerical study \cite{KeLa93}.

Recently Salmhofer \cite{Salmhofer} has proved the existence of a
unique density of zeroes in the thermodynamic ($N\rightarrow \infty$)
limit.  The scaling behaviour of the Yang--Lee edge in the
thermodynamic limit was studied by Abe \cite{AbeLY} and by Suzuki
\cite{SuzukiLY} in 1967 for Ising models below the upper critical
dimension. They found asymptotic forms for the density of zeroes and a
power--law behaviour for the scaling of the edge in the symmetric
phase.

Here we would like to add to this body of knowledge by presenting some
results on Lee--yang zeroes in the symmetric phase ($t>0$) of the four
dimensional Ising model. To this end we find an asymptotic form for the
density of zeroes in the thermodynamic limit which is sufficient to
recover the scaling forms for the specific heat and susceptibility.
The finite--size behaviour of the location and density of the Lee--Yang
zeroes is studied both analytically and  numerically.

\section{ The Density of Lee--Yang Zeroes and the Yang--Lee Edge}
\setcounter{equation}{0}

According to the Lee--Yang theorem \cite{LY} the zeroes of the partition
function all lie on the unit circle in the complex fugacity plane.
Denoting the ($t$--dependent) position of these zeroes by
\begin{equation}
 z_j (t) = e^{i\theta_j(t)}\quad ,\quad \theta_j\in\Re
\quad ,\quad j = 1,\dots, N
\end{equation}
the partition function may be written as
\begin{equation}
 Z_N(t,z) = z^{-\frac{N}{2}}
 \rho_N(t)
 \prod_{i=1}^{N}
 \left( z - e^{i\theta_j(t)}
 \right)
\quad .
\label{second}
\end{equation}
The largest coefficient $\rho_N(t)$ plays no r\^ole in the following and
we henceforth set it to unity. The free energy density,
\begin{equation}
 f_N(t,z)
 =
 \frac{1}{N}
 \ln{Z_N(t,z)}
\quad ,
\end{equation}
can be written as
\begin{equation}
 f_N(t,z)
 =
 -\frac{1}{2}
 \ln{z}
 +
 \frac{1}{N}
 \sum_{j=1}^{N}
 \ln{\left(
 z - e^{i\theta_j(t)}
 \right)}
\quad .
\end{equation}
The discrete measure $dG_N$ is formally given by
\begin{equation}
 g_N(\theta,t)
 =
 \frac{dG_N(\theta,t)}{d\theta}
 =
 \frac{1}{N}
 \sum_{j=1}^{N}{\delta(\theta - \theta_j(t))}
\quad .
\end{equation}
The $t$--dependent density of Lee--Yang zeroes on the unit circle in
the complex $z$ plane is given by $g_N$ and the cumulative density of
zeroes $G_N$ is a function monotonically increasing in $\theta$ from
$G(0,t)=0$ to $G(2\pi,t)=1$. The free energy is
\begin{equation}
 f_N(t,z) =
 -\frac{1}{2} \ln{z}
 +
 \int_{\theta =0}^{\theta = 2 \pi}{
 \ln{\left( z - e^{i \theta} \right)}
 d G_N(\theta,t)
 }
\quad .
\label{old2.6}
\end{equation}
The thermodynamic limit is
\begin{eqnarray}\label{fxx}
 g(\theta,t) & = & \lim_{N\rightarrow \infty} g_N(\theta,t) \quad , \\
 G(\theta,t) & = & \lim_{N\rightarrow \infty} G_N(\theta,t)\quad , \\
 f(t,z) & = & \lim_{N\rightarrow \infty} f_N(t,z)\quad .
\end{eqnarray}
The coefficients $\rho_k(t)$ of the polynomial (\ref{polynomial}) are real
and hence $g(-\theta,t)=g(\theta,t)$.
Therefore it is sufficient to consider only the interval $0\leq
\theta\leq\pi$ in the integrals.
The Yang--Lee edge $\theta_c(t)$ is defined by
\begin{equation}
 g(\theta,t)=0 {\rm{~ ~ for ~ ~}} -\theta_c(t) < \theta < \theta_c(t)
\quad .
\end{equation}
Integrating (\ref{old2.6}) by parts gives for the free energy
\begin{equation}
 f(t,z)
 =
 \frac{1}{2}
 \ln{(2\cosh{(2h)} + 2)}
 -
 \int_{\theta_c(t)}^{\pi}
 \frac{\sin{\theta}}{\cosh{(2h)}-\cos{\theta}}
 G(\theta,t)
 d\theta
\quad .
\label{free1}
\end{equation}
The magnetization is then
\begin{equation}
 \frac{\partial f}{\partial h}
 =
 \tanh{(h)}
 +
 2 \sinh{(2h)}
 \int_{\theta_c(t)}^{\pi}
 \frac{\sin{\theta}}{(\cosh{(2h)}-\cos{\theta})^2}
 G(\theta,t)
 d \theta
\quad ,
\end{equation}
and the zero field susceptibility
\begin{equation}
 \chi(t)
 = \left( \frac{\partial^2 f}{\partial h^2}\right)_{h=0}=
 1  + 4  \int_{\theta_c(t)}^{\pi}
 \frac{\sin{\theta}}{(1-\cos{\theta})^2}
 G(\theta,t)
 d \theta
\quad .
\label{zerofieldX}
\end{equation}
One expects the contribution of small $\theta$ to be dominant
\cite{AbeLY,SuzukiLY}. In particular we want to study its
contribution singular in $t$. Expanding the trigonometric functions
in (\ref{zerofieldX}) (and dropping the constant term),
\begin{equation}
 \chi(t)
 =
 16
 \int_{\theta_c(t)}^{\pi}
 \frac{G(\theta,t)}{\theta^3}
 \{1 + O(\theta^2) \}
 d \theta
\quad .
\label{equat26}
\end{equation}
In four dimensions and in the symmetric phase the perturbative
renormalization group gives \cite{BLZ}
\begin{equation}
 \chi \left( t \right)
 \sim
 t^{-1} \left( -\ln{t} \right)^{\frac{1}{3}}
 \quad .
\label{myAbe72-3.1}
\end{equation}
A change of variables is introduced via
$\theta = \theta_c x$. Then in the critical region where $t>0$ is sufficiently
small
\begin{equation}
 t^{-1}
 ( -\ln{t} )^{1/3}
 \sim
 {\theta_c(t)}^{-2}
 \int_1^{\pi / \theta_c(t)}
 \frac{G(x\theta_c ,t)}{x^2}dx
 \quad .
\end{equation}
Following \cite{AbeLY,SuzukiLY}, the upper integral limit can be replaced by
infinity near criticality. This leads to the requirement that
\begin{equation}
 \frac{t (-\ln{t})^{-\frac{1}{3}}}{\theta_c(t)^2}
 \int_1^\infty
 \frac{G(x\theta_c,t)}{x^2}
 dx
 \sim
 {\rm{constant.}}
\end{equation}
For fixed $t$ the integral is bounded due to the boundedness of $G$.
The constancy  leads to a differential
equation \cite{AbeLY,SuzukiLY}
for $G$ with the general solution
\begin{equation}
 G(\theta,t)
 =
 t^{-1}
 \left( -\ln{t} \right)^{ \frac{1}{3} }
 { \theta_c(t) }^2
 \Phi \left( \frac{\theta}{\theta_c(t)} \right)
\quad ,
\label{twoPhi}
\end{equation}
$\Phi (x)$ being an arbitrary function of $x$ with
$\Phi(\mid x\mid\leq 1)=0$. Then
\begin{equation}
 g(\theta,t)
 =
 \frac{d G(\theta,t)}{d \theta}
 =
 t^{-1}
 \left( -\ln{t} \right)^{ \frac{1}{3} }
 \theta_c(t)
 \Phi^\prime \left( \frac{\theta}{\theta_c(t)} \right)
\label{myAbe3.4}
\end{equation}
where $\Phi^\prime (x) = \frac{d \Phi (x)}{dx}$.

{}From (\ref{free1}) (and using the fact that
$G\left(\theta_c,t\right)=0$), one gets the specific heat
\begin{equation}
 C_V(t) =  \left.
 \frac{\partial^2f(t,z)}{\partial t^2}
 \right|_{h=0} =
 -2\int_{\theta_c(t)}^\pi
 \theta^{-1}
 \frac{d^2G\left(\theta,t \right)}{dt^2}
 \{1 + O(\theta^2)\}d\theta
 \quad .
\label{directerthanAbe}
\end{equation}
Now the cumulative density of zeroes in four dimensions may be found
from (\ref{twoPhi}).  In four dimensions one expects the power law
scaling behaviour characteristic of dimensions below the upper critical
one to be modified by multiplicative logarithmic corrections. Assume
therefore that the Yang--Lee edge has the scaling behaviour
\begin{equation}
 \theta_c(t) = At^p\left(-\ln{t}\right)^{-\lambda}
\label{assumedLYedge}
\end{equation}
for small $t>0$ and with $0<p<1$. This gives
\begin{eqnarray}
\lefteqn{\frac{d^2G\left(\theta,t\right)}{dt^2}
 = A^2
 t^{2p-3}\left( -\ln{t}\right)^{\frac{1}{3}-2\lambda}
 \left[1+ O\left( \frac{1}{\ln{t}} \right) \right]
}
\nonumber \\
 & &
 \times
 \left\{ 2(1-3p+2p^2)\Phi(x) + p(3-4p)x\Phi^\prime(x)
 + p^2x^2\Phi^{\prime\prime}(x)
 \right\}
\end{eqnarray}
where $x= \theta / \theta_c$ and a prime indicates derivative with
respect to $x$. The specific heat is then
\begin{equation}
 C_V \propto
 t^{2p-3}\left( -\ln{t}\right)^{\frac{1}{3}-2\lambda}
 \left[1+ O\left( \frac{1}{\ln{t}} \right) \right]
 \int_1^{\frac{\pi}{\theta_c}}
 {\frac{I(x)}{x} dx}
 \quad ,
\label{myAbe72-3.20a}
\end{equation}
where $I$ is some function of $x$.
As $t\rightarrow 0$ ($\theta_c(t) \rightarrow 0$),
one has
\begin{equation}
 C_V \propto t^{2p-3}\left( -\ln{t} \right)^{\frac{1}{3}-2\lambda}
 \left[1+ O\left( \frac{1}{\ln{t}} \right) \right]
\label{directC}
\end{equation}
in the symmetric phase ($t>0$, $H=0$) and near criticality.  From
perturbation renormalization group analyses it is known \cite{BLZ} that
the zero field specific heat scales as
\begin{equation}
 C_V(t) \sim \left(-\ln{t} \right)^{\frac{1}{3}}
\label{Cv4DfromBLZ}
\end{equation}
in four dimensions. Therefore $p=\frac{3}{2}$ and $\lambda=0$.
 From (\ref{assumedLYedge}) the Yang--Lee edge in four dimensions
scales as
\begin{equation}
 \theta_c(t) \sim t^{\frac{3}{2}}
 \quad .
\label{myAbe3.9}
\end{equation}
This is the same formula as that yielded by mean field theory
\cite{IPZ,AbeLY,SuzukiLY}.

The density of Lee--Yang zeroes is given by (\ref{myAbe3.4}) as
\begin{equation}
 g(\theta,t) =t^{\frac{1}{2}}\left( -\ln{t}\right)^\frac{1}{3}
 \Phi^\prime \left( \frac{\theta}{\theta_c} \right) \quad ,
\label{densityofLYzeroes} \end{equation} in which $\Phi^\prime$ is an
unknown function.  This form is sufficient to recover the singular
behaviour of the susceptibility and of the specific heat.

The behaviour of the zero at $\theta = x \theta_c$ (for fixed $x$) as a
function of $t$ ($t>0$) is given by (\ref{densityofLYzeroes}). At fixed
$t$, $g(\theta,t)$ is an unknown function of $\theta/\theta_c$.
Kortman and Griffiths emphasized the study of the density of zeroes
close to the Yang--Lee edge \cite{KoGr71}. Using high temperature and
high field series they concluded that  below the upper critical
dimension and  for  a fixed  (strictly positive) $t$, the density of
zeroes near the edge exhibits a power law behaviour
\begin{equation}
 g(\theta,t)
 \sim
 \left(\theta - \theta_c(t)
    \right)^\sigma
\quad .
\label{sig}
\end{equation}

In zero  dimensions (a single site) $\sigma$ is known to be $-1$
\cite{BaIt84}. For the exactly solvable one dimensional Ising model
$\sigma = -1/2$ for all $t>0$ \cite{LY}, i.e., the density of zeroes
diverges as the edge is approached.   The Ising model in the presence
of an external field has not   been solved   in more than one
dimension. Nonetheless the value of $\sigma$ in two dimensions has been
found to be $-1/6$ by Dhar \cite{Dh83} by mapping the two dimensional
Ising ferromagnet into a solvable model of three dimensional directed
animals. Cardy \cite{Ca85} found the same result by using the conformal
invariance of two dimensional systems at the  critical point.  Using
high temperature numerical methods, Kurtze and Fisher
\cite{Fi78,KuFi79} found  $\sigma  =  0.086(15)$ in three dimensions.
It is believed that this values hold independent of the lattice
parametrization used \cite{Fi78}. For the mean field theory $\sigma =
1/2$ \cite{KoGr71}. Thus there seems to   be a systematic increase of
$\sigma$ with dimensionality.

At criticality $t=0$, however, the Yang--Lee gap vanishes
and one may expect the critical exponent $\sigma$ to take
on a value different than that in the symmetric phase.
Now, the density of zeroes is proportional to the discontinuity in the
magnetization M crossing the locus of zeroes \cite{LY}
\begin{equation}
 \lim_{r\rightarrow 1^+}{M(t,z=re^{i\theta})}
 -
 \lim_{r\rightarrow 1^-}{M(t,z=re^{i\theta})}
  \propto
 g(\theta,t)
\quad .
\end{equation}
The infinite volume behaviour of the magnetization below the upper critical
dimension
\begin{equation}
 M(t=0,H) \sim H^{\frac{1}{\delta}}
\label{bbb}
\end{equation}
should be recovered from (\ref{sig}) at $t=0$ and therefore, for $d<4$,
\begin{equation}
 g(\theta,t=0) \sim \theta^{\frac{1}{\delta}}
\quad .
\label{delta}
\end{equation}
In four dimensions where $\delta = 3$, one expects the above formulae  to be
modified by multiplicative logarithmic corrections.
There, (\ref{bbb}) becomes \cite{BLZ}
\begin{equation}
 M(t=0,H) \sim H^{\frac{1}{3}}(-\ln{H})^{\frac{1}{3}}
 \quad .
\end{equation}
Therefore, in 4D in the thermodynamic limit
\begin{equation}
 g(\theta,t=0) \sim \theta^{\frac{1}{3}}(-\ln{\theta})^{\frac{1}{3}}
\quad .
\end{equation}

\section{ Finite--Size Analysis}
\setcounter{equation}{0}

Non--perturbative means of calculating thermodynamic functions in spin
models are provided by stochastic techniques like Monte Carlo
integration.  These numerical methods yield exact results subject only
to statistical error.  They are however limited to finite lattices. One
has to rely on finite--size scaling (FSS) extrapolation methods to gain
information on the corresponding thermodynamic limit.

Let $P_L(t)$ represent the value of some thermodynamic quantity $P$ at
reduced temperature $t$ on a lattice characterized by a linear extent $L$.
Then, if $\xi $ is the correlation length, the FSS hypothesis is that
\cite{KeLa93,Ba83}
\begin{equation}
 \frac{P_L(t)}{P_\infty(t)} = f\left( \frac{\xi_L(t)}{\xi_\infty(t)}
 \right)
 \quad .
\label{FSShyp}
\end{equation}
In four dimensions the scaling behaviour of the correlation length is
\cite{LW}
\begin{equation}
 \xi_\infty (t) \sim t^{-\frac{1}{2}} (-\ln{t})^{\frac{1}{6}}
\quad .
\end{equation}
Its scaling with $L$ is \cite{Br82}
\begin{equation}
 \xi_l (0) \sim L (\ln{L})^{\frac{1}{4}}
\quad .
\end{equation}
Therefore in $d=4$ the scaling variable should include
logarithmic terms \cite{KeLa93},
\begin{equation}
 x
 =
 \frac{
 L ( \ln{L} )^{ \frac{1}{4} }
 }{
 t^{ -\frac{1}{2} }
 (-\ln{t})^{ \frac{1}{6} }
 }
\quad .
\end{equation}
Let $H_1$ be the position of the Yang--Lee edge in the complex external
magnetic field plane in the thermodynamic limit and let $H_1(L)$ be its
finite--size counterpart (i.e., the position of the lowest lying zero
for a system of finite linear extent $L$). From (\ref{myAbe3.9}),
the FSS hypothesis applied to the Yang--Lee edge gives
\begin{equation}
 H_1(L) \sim t^{\frac{3}{2}} f(x)
\quad .
\end{equation}
Fixing $x$ (so that when rescaling $L$, the temperature is also
rescaled in such a way as to keep $x$ constant), we find
\begin{eqnarray}
 H_1(L)
 & \sim &
 \left(
 x^{-1} L^{-2}
 (\ln{L})^{-\frac{1}{6}}
 \right)^{\frac{3}{2}}
 f(x) \nonumber \\
 ~ & \sim &
 L^{-3}
 (\ln{L})^{-\frac{1}{4}}
\quad .
\label{toget}
\end{eqnarray}
This FSS formula agrees with that derived recently by perturbative
renormalization
group methods \cite{KeLa93}.

The perturbative RG analysis of the finite--size $\phi^4_4$ model
\cite{KeLa93} gives the relationship between the magnetization $M_L(t,H)$
and external field $H$ at reduced temperature $t$
\begin{equation}
 H =  c_1t M_L (\ln{L})^{-\frac{1}{3}}
      + c_2 M_L^3 (\ln{L})^{-1}
\quad ,
\end{equation}
where $c_1$ and $c_2$ are constants. At $t=0$, therefore,
\begin{equation}
 M_L(t=0,H) \sim H^{\frac{1}{3}} (\ln{L})^{\frac{1}{3}}
\quad .
\end{equation}
For a finite--size system the position of the Yang--Lee edge is not
zero at $t=0$ and the origin of non--vanishing density of zeroes has to
be correspondingly shifted as in (\ref{sig}). One therefore expects
the  density of zeroes to be
\begin{equation}
 g_L(H_j(L)) \sim (H_j(L)-H_1(L))^{\frac{1}{3}} (\ln{L})^{\frac{1}{3}}
\quad ,
\label{smllg}
\end{equation}
where $H_j(L)$ is the position of the $j^{\rm{th}}$ Lee--Yang zero.
Defining the cumulative density of zeroes at the $j^{\rm{th}}$
zero by the fractional total of zeroes up to $H_j(L)$,
\begin{equation}
 G_L(H_j(L)) = \frac{j-1}{L^4}
\quad ,
\label{bigg}
\end{equation}
we find (integrating $g_L$ in (\ref{smllg}) to $G_L$)
\begin{equation}
 \frac{j-1}{L^4} \sim (H_j(L)-H_1(L))^{\frac{4}{3}}
(\ln{L})^{\frac{1}{3}}
\quad .
\label{311}
\end{equation}
Therefore
\begin{equation}
 H_j(L)-H_1(L)
 \sim
 \left(\frac{j-1}{L^4}\right)^{\frac{3}{4}}(\ln{L})^{-\frac{1}{4}}
\quad .
\label{recover}
\end{equation}
Eq.(\ref{recover}) gives the $j$ dependence of the lowest lying zeroes
as well as recovering the FSS prediction of (\ref{toget}).

We now compare these FSS results with data obtained for the 4D Ising
model in a high statistics Monte Carlo calculation.  The simulation was
done with the Swendsen--Wang cluster updating algorithm \cite{SwWa87}
applied to lattices of sizes $L^4$ for lattice sizes with linear
extension $L=8,12,16,20,24$ (details of the numerics can be found in
\cite{KeLa93}).

Histogram approximations to the spectral density $\rho (S,M)$ of
(\ref{gndcanIsing1}) were determined at zero external field and at
various values of $\kappa$ close to the pseudocritical value (the value
of $\kappa$ at which the zero field specific heat peaks). For the
histogram in the magnetization each of the raw histograms in $S$ and
$M$ were firstly binned in a $256\times256$ array. For each $M$--bin
the corresponding $S$--subhistograms were then combined to
multi-histograms according to \cite{FeSw88}.  In this way, an optimal
histogram in $M$ for arbitrary $\kappa$ was obtained. From this the
partition function may be determined for not too large values of
(complex) $H$.

The critical value of $\kappa$ in four dimensions has been determined
to $\kappa_c = 0.149703(15)$ \cite{KeLa93}.  Our data yield only three
reliable Lee--Yang zeroes for each lattice size. The reason for this is
demonstrated in fig.1 where the contours along which ${\rm{Re}}Z=0$ and
${\rm{Im}}Z=0$ (for $L=24$ and $\kappa = 0.149703$) are plotted.
Because of the magnification of statistical errors far away from the
simulation point $H=0$ these contours fail to cross the imaginary $Z$
axis when ${\rm{Im}}H$ is large. Thus the zeroes move off the
${\rm{Im}}H$ axis and their positions are unreliable. The remaining
lattices give qualitatively similar pictures.

Table 1 lists the positions of the first Lee--Yang zeroes (the
Lee--Yang edge) as obtained from the multihistograms for various
$\kappa$ values near $\kappa_c$ and for all five lattices analyzed.
 As $\kappa$ increases one expects the zeroes to approach the real axis
in the thermodynamic limit according to (\ref{myAbe3.9}). Fig.2 shows
the corresponding behaviour for the finite--size systems considered. At
$\kappa_c$ they should scale according to the FSS formula
(\ref{toget}).

Table 2 lists the positions of the first three Lee--Yang zeroes as
obtained from the multihistogram at our estimated value for the
critical coupling in the infinite volume limit, $\kappa_c = 0.149703$.
The errors in the quantities calculated from the multihistograms were
estimated by the jackknife method. The data for each lattice size were
cut to produce $10$ subsamples leading to $10$ different
multihistograms.  These $10$ different multihistograms give $10$
different results for the quantities in question, whence the variance
and bias were calculated.

The density of zeroes should behave according to  (\ref{smllg}) or
(equivalently) (\ref{recover}).  The log--log plot of fig.3
 gives a slope of $0.778(2)$.  The deviation from the exponent $0.75$
in (\ref{recover}) is presumably due to the presence of logarithmic
corrections. This may be seen in fig.4 where we remove the expected
leading behaviour: A negative slope is clearly identified. In fact a
best fit to all ten points gives a slope $-0.248(17)$. The shaded area
is bordered by lines of this slope.

We find that both  leading power--law scaling behaviour and
multiplicative logarithmic corrections for the density of zeroes (or
equivalently for the distance between zeroes) are identified in figs. 3
and 4. This is complementary to our previous analysis in which the
scaling behaviour of the actual positions of these zeroes was analyzed
\cite{KeLa93}.  Both approaches yield quantitative agreement with the
(perturbative) theoretical predictions.

\section{Conclusions}
\setcounter{equation}{0}

The scaling behaviour of the Lee--Yang zeroes and in particular of the
Yang--Lee edge in four dimensions and in the thermodynamic limit has
been examined. The asymptotic form for the density of zeroes in the
infinite volume limit is sufficient to recover the scaling formulae for
the specific heat, the magnetization and the magnetic susceptibility.
This extends the work of Abe and Suzuki to the case of four dimensions
where mean field power--law scaling behaviour is modified by
multiplicative logarithmic corrections which are linked to the
triviality of the theory.  An analytical FSS study of the edge and the
density of zeroes is in good quantitative agreement with  a numerical
analysis in the form of Monte Carlo simulations on finite size
lattices.

\newpage


\newpage

\section*{Tables}
\setcounter{equation}{0}

{\noindent \bf Table 1:} The positions of the first Lee--Yang zeroes as
obtained from the multihistograms for all five lattices and near
$\kappa_c$.  The real part of the zeroes is always zero.

\smallskip
\begin{center}
\begin{tabular}{|c|c|c|c|c|c|}
\hline
 & $L=8$ & $L=12$ & $L=16$ & $L=20$
 & $L=24$ \cr
 $\kappa$ & $\mbox{Im} H_{1}$ & $\mbox{Im} H_{1}$
 & $\mbox{Im} H_{1}$ & $\mbox{Im} H_{1}$ & $\mbox{Im} H_{1}$ \cr
\hline
 0.149600 & 0.015281 & 0.004511 & 0.001958 & 0.001047 & 0.000637
\cr
 0.149650 & 0.015091 & 0.004384 & 0.001860 & 0.000966 & 0.000567
\cr
 0.149703 & 0.014892 & 0.004253 & 0.001761 & 0.000886 & 0.000500
\cr
 0.149750 & 0.014718 & 0.004140 & 0.001677 & 0.000820 & 0.000447
\cr
 0.149800 & 0.014535 & 0.004023 & 0.001593 & 0.000756 & 0.000398
\cr
 0.149850 & 0.014355 & 0.003910 & 0.001512 & 0.000697 & 0.000355
\cr
 0.149900 & 0.014177 & 0.003800 & 0.001437 & 0.000644 & 0.000319
\cr
 0.149950 & 0.014001 & 0.003693 & 0.001366 & 0.000597 & 0.000289
\cr
 0.150000 & 0.013828 & 0.003590 & 0.001299 & 0.000555 & 0.000264
\cr
 0.150050 & 0.013657 & 0.003491 & 0.001237 & 0.000518 & 0.000244
\cr
 0.150100 & 0.013488 & 0.003395 & 0.001180 & 0.000486 & 0.000228
\cr
 0.150150 & 0.013322 & 0.003302 & 0.001127 & 0.000458 & 0.000215
\cr
 0.150200 & 0.013159 & 0.003213 & 0.001077 & 0.000433 & 0.000204
\cr
 0.150250 & 0.012997 & 0.003127 & 0.001032 & 0.000413 & 0.000195
\cr
 0.150300 & 0.012838 & 0.003044 & 0.009909 & 0.000395 & 0.000187
\cr
 0.150350 & 0.012682 & 0.002965 & 0.009530 & 0.000379 & 0.000180
\cr
 0.150400 & 0.012527 & 0.002889 & 0.009185 & 0.000366 & 0.000174
\cr
\hline
\end{tabular}
\end{center}

\smallskip

{\noindent \bf Table 2:} The positions of the first three Lee--Yang
zeroes as obtained from the jackknifed multihistograms at $\kappa =
0.149703$.  The real part of the zeroes is always zero.

\vspace{0.5cm}

\begin{center}
\begin{tabular}{|r|l|l|l|}
\hline
 $L$ & $\mbox{Im}(H_{1})$ & $\mbox{Im}(H_{2})$ &
$\mbox{Im}(H_{3})$ \cr
\hline
 8 & 0.014892(22) & 0.033057(48) & 0.047357(174) \cr
 12 & 0.004253(16) & 0.009426(15) & 0.013349(71) \cr
 16 & 0.001761(6) & 0.003905(22) & 0.005388(24) \cr
 20 & 0.000886(5) & 0.001970(12) & 0.002743(29) \cr
 24 & 0.000500(4) & 0.001106(5) & 0.001541(12) \cr
\hline
\end{tabular}
\end{center}

\newpage

\section*{Figures}

{\noindent \bf Fig 1:}
Contours along which ${\rm{Re}}Z=0$ (dotted lines) and ${\rm{Im}}Z=0$
(full lines) (for $L=24$ and $\kappa =0.149703$).
{}~\\

{\noindent \bf Fig 2:}
The zeroes approach the real axis as $\kappa$ increases;
at $\kappa_c$ they should scale with the lattice size $L$
according to (\ref{toget}). Here the triangles, circles,
diamonds, stars and crosses correspond to lattice sizes 8,12,16,20
and 24 respectively.
{}~\\

{\noindent \bf Fig 3:}
The FSS of the density of zeroes is given by (\ref{recover}).
The leading power--law behaviour is revealed by a log--log plot.
Here the open  diamonds and triangles correspond to $j=2$ and $j=3$
respectively.
This gives a slope 0.778(2), the deviation away from 0.75
being due to the presence of logarithmic corrections.
{}~\\

{\noindent \bf Fig 4:}
Data like in fig.3, but with the leading power--law behaviour
removed; we clearly identify the negative exponent in the $\log L$
behaviour. The shaded band indicates the result of a fit giving a
slope value of $-0.248(17)$.
\end{document}